\documentclass{aa}  
\usepackage{gensymb}
\usepackage{hyperref}
\hypersetup{
    colorlinks = true, 
    urlcolor = cyan, 
    linkcolor = blue, 
    citecolor = blue}

\usepackage{graphicx}
\usepackage{txfonts}
\usepackage{multirow}
\usepackage{xcolor}
\usepackage{booktabs}

\begin{document}

   \title{Exploring the optical properties of redback pulsars: The case of J1717$+$4308A in the globular cluster M92}

   \author{Jianxing Chen
          \inst{1,2,3}
          \and
          Greta Ettorre
          \inst{2,3}
          \and
          Cristina Pallanca
          \inst{2,3}
          \and
          Mario Cadelano
          \inst{2,3}
          \and
          Bidisha Sen
          \inst{4}
          \and 
          Devina Misra
          \inst{4} 
          \and
          Emanuele Dalessandro
          \inst{3}
          \and
          Francesco R. Ferraro
          \inst{2,3}
          \and 
          Barbara Lanzoni
          \inst{2,3}
          \and
          Alessandro Ridolfi
          \inst{5,6}
          \and
          Marta Burgay
          \inst{5}
          \and
          Andrea Possenti
          \inst{5,7}
          \and
          Paulo C. C. Freire
          \inst{8}
          \and
          Scott M. Ransom
          \inst{9}
          \and 
          Manuel Linares
          \inst{4,10}
          \and
          Rene P. Breton 
          \inst{11}
          }

   \institute{School of Physics and Astronomy, Beijing Normal University, No.19, Xinjiekouwai St, Haidian District, Beijing, 100875, People's Republic of China
   \and
             Dipartimento di Fisica e Astronomia ``Augusto Righi'', Alma Mater Studiorum Universit\`a di Bologna, via Piero Gobetti 93/2, I-40129 Bologna, Italy
    \and
             INAF-Osservatorio di Astrofisica e Scienze dello Spazio di Bologna, Via Piero Gobetti 93/3 I-40129 Bologna, Italy
    \and
             Department of Physics, Norwegian University of Science and Technology, NO-7491 Trondheim, Norway
    \and
            INAF - Osservatorio Astronomico di Cagliari, Via della Scienza 5, 09047, Selargius, CA, Italy 
    \and
            Max-Planck-Institut für Radioastronomie, Auf dem Hügel 69, 53121, Bonn, Germany
    \and
            Università di Cagliari, Dipartimento di Fisica, S.P. Monserrato-Sestu Km 0,700, 09042, Monserrato, CA, Italy
    \and
            Max-Planck-Institut für Radioastronomie, Auf dem Hügel 69, 53121, Bonn, Germany
    \and
            National Radio Astronomy Observatory (NRAO), Charlottesville, VA 22903, USA
    \and
            Departament de Física, EEBE, Universitat Politècnica de Catalunya, Av. Eduard Maristany 16, E-08019 Barcelona, Spain
    \and
            Jodrell Bank Centre for Astrophysics, Department of Physics and Astronomy, The University of Manchester, Manchester M13 9PL, UK
    \and
            School of Physics, University College Cork, Cork, Ireland
      }

   \date{Received xxyyzz; accepted xxyyzz}

  \abstract
 {Binary millisecond pulsars (MSPs) in globular clusters (GCs) are key for binary and stellar evolution studies under extreme conditions. The identification of their optical companion stars is instrumental in order to characterise these systems and to constrain the possible recycling mechanisms. 
 For this work, we searched for the optical counterpart to PSR J1717+4308A (hereafter M92A) in the GC M92. 
To this end, we exploited a multi-epoch, multi-wavelength dataset obtained with the Hubble Space Telescope. 
We constructed colour--magnitude diagrams, investigated proper motions to assess cluster membership, and modelled the observed light curves. We identified an object located at only 0.02 arcsec from the nominal radio position as the likely optical companion to M92A. The star is significantly bluer than the main sequence at the same luminosity level and exhibits clear photometric variability  with a periodicity in agreement with the orbital motion of the binary. The light curve displays two maxima and two minima, indicative of strong tidal distortion and only mild irradiation. 
Such mild irradiation is consistent with the ratio of the pulsar spin-down to the companion flux ($f_\mathrm{sd}$), which for M92A lies close to the boundary between ellipsoidal- and irradiation-dominated regimes ($f_\mathrm{sd} \approx 2.71$).
From the light curve modelling we inferred the main physical properties of the companion star. The best-fit model indicates a high-inclination system with a relatively low-mass companion and a massive neutron star. With a base temperature of $\sim7200$ K, the companion ranks among the hottest redbacks known to date. 
This object therefore adds additional pieces to the puzzle of MSP companion properties and contributes to outlining the characteristics of redbacks across the different classes.
}

   \keywords{Millisecond pulsars --
                Binary pulsars --
                Globular clusters
               }
               
   \titlerunning{Redback millisecond pulsar M92A}
   \maketitle

\section{Introduction}
\label{sec-1}
With spin periods below 30 ms, millisecond pulsars (MSPs) represent the most rapidly rotating class of neutron stars (NSs). They are recycled systems, where an old and slowly rotating NS has been spun up through mass accretion and angular momentum transfer from a companion star 
(e.g. \citealt{Alpar+82, Bhattacharya+91, Wijnands+98, D'Antona+22, Tauris+23}).

About 600 MSPs have been discovered in the Galaxy to date,\footnote{The ATNF pulsar catalogue: \url{https://www.atnf.csiro.au/research/pulsar/psrcat/}} and half of them are found within globular clusters (GCs).\footnote{Pulsars in GCs: \url{http://www.naic.edu/~pfreire/GCpsr.html}} 
While this ratio is partially due to an observational bias, given that GCs have long been and are currently targeted for deep pulsar search observations, it also reflects the impact of the environment in the formation of these objects. In fact, GCs are ideal MSP factories due to their high stellar densities, which create collisional environments where dynamical interactions among stars and binaries favour the production of a large number 
of exotic objects \citep{ivanova06_wd,ivanova08_ns}, such as MSPs 
\citep{Cadelano+17, Cadelano+18, Cadelano+20, Ridolfi+22, Yin+24}, blue straggler stars \citep{Ferraro+93, Ferraro+97, Ferraro+99, Ferraro+01a, Ferraro+03a, Ferraro+12, Ferraro+18, Cadelano+22b, Ferraro+23, Ferraro+26,Dalessandro+08}, cataclysmic variables \citep{paresce+92,Pooley+06,Knigge+12,Bao+23}, and compact object binaries \citep{Barr+24}. 

Millisecond pulsars are generally thought to be the final outcome of the evolution of 
low-mass X-ray binaries (LMXBs), where the NS recycling process is directly observed via intense X-ray emission as a result of the ongoing accretion. At the end of the accretion phase, the companion is expected to be the stripped core of the original donor star, typically a He white dwarf (WD)
\citep{Ferraro+03a,Rivera-Sandoval+15,Cadelano+15b, Cadelano+19, Cadelano+20, Chen+23, Ettorre+25a, Lian+25a, Li+25}. However, the evolutionary pathways leading to MSP formation can also give rise to distinct sub-classes of binary systems, the most notable of which are  called  spiders, which are tight binaries with periods shorter than $\sim1$ day, usually characterised by periodical eclipses of the pulsar signal due to interposed ionised material, likely caused by the ongoing ablation of a non-degenerate companion star by the pulsar relativistic wind. The spider pulsar class is divided into two sub-classes: black widows and redbacks. The former have very low-mass companions ($\mathrm{M_{COM}}<0.05 \,\mathrm{M}_{\odot}$), while the latter 
have  more massive companions ($\mathrm{M_{COM}}\sim0.1-0.5 \,\mathrm{M}_{\odot}$). In the optical bands, spider companions usually show magnitude modulation due to tidal distortion and irradiation of the stellar side exposed to the pulsar flux \citep{Roberts+13, Swihart+22, Turchetta+23}. 
Subsequent works have expanded the sample of optically identified spider 
companions \citep[e.g.][]{Bobakov+24, Bobakov+25}, modelled their light 
curves with \textsc{Icarus} \citep{Kandel+20, Kandel+23, Mata+23}, 
and performed systematic searches for new systems \citep{Turchetta+26}.

The physical mechanism producing redbacks and black widows is still debated (see e.g. \citealt{D'Antona+22}). \citet{Benvenuto+12, Benvenuto+13, Benvenuto+14}, and \citet{Misra+25} modelled the evolution of NS binaries and argued that redbacks represent an early stage of evolution towards the black widow late stage, although some redback companions might evolve to canonical He-core WD systems (see \citealt{Burderi+02}). Conversely, \citet{Chen+13} suggested that black widows and redbacks are two independent outcomes of the LMXB evolution, and that the reprocessing efficiency of the pulsar luminosity by the companion star is the leading term in generating a redback or a black widow. In this respect, it is interesting to mention that the link between LMXBs and MSPs (and, in particular, redbacks) has been tightened with the discovery of the transitional MSPs \citep{Papitto+13, Papitto+20}. These are sources that rapidly switch between an accretion-powered regime and a rotation-powered regime. During the former regime, mass accretion is active and the pulsar is not  
visible in the radio bands, as typically observed in LMXBs. In the latter regime, the accretion is halted and the pulsar is visible again in the radio bands as an eclipsing MSP, whose properties closely resemble those of redbacks.

The identification of the optical companion stars to MSPs is the key to characterising these systems and constraining the possible recycling and formation mechanisms. 
In this context, we are conducting a long-term programme focused on identifying and characterising the optical counterparts to MSPs in GCs. To date, a large number of He-core WD companions have been discovered \citep{Ferraro+03a, Cadelano+15b, Cadelano+19, Cadelano+20, Chen+23, Ettorre+25a}, along with companions to spider MSPs \citep{Pallanca+10, Pallanca+13, Pallanca+14, Cadelano+15a}.
Here we present the identification and optical characterisation of the 
recently
discovered redback system PSR~J1717+4308A (hereafter M92A) located in the GC NGC 6341 (M92). M92 is a metal-poor GC with $\mathrm{[Fe/H]} = -2.31 $ \citep[][2010 Edition]{Harris+96}, 
an age of $\sim 13$ Gyr \citep{Dotter+10}, and a low extinction of $\mathrm{E(B-V)} = 0.02$. M92A was first reported as an eclipsing binary MSP by \citet{Pan+20} based on observations with the Five-hundred-meter Aperture Spherical radio Telescope (FAST, \citealt{Nan+11}). It has a spin period of 3.16 ms and a binary orbital period of 0.2 days. \citet{Pan+20} also estimated that the companion should be a low-mass main-sequence or sub-giant star with a median mass of 0.18 $\mathrm{M}_{\odot}$, suggesting that M92A is a redback system. 
The nominal radio position of M92A at the reference epoch MJD 58390.00 is $\alpha_{J2000}=17^\mathrm{h}\!:17^\mathrm{m}\!:06^\mathrm{s}\!.49640$ and $\delta_{J2000}=+43\degree\!:08'\!:03''\!.4786$ \citep[from the FAST Globular Cluster Pulsar Survey, GC FANS;][]{Lian+25b}.
 Moreover, a $\gamma$-ray emission associated with M92A was detected by \citet{Zhang+23} at a $4.4 \sigma$ confidence level, with a luminosity of $1.3 \times 10^{34}$ erg s$^{-1}$. 
The X-ray counterpart was discovered by \citet{Zhao+22}, although the source was already identified by \citet{Lu+11}. The X-ray spectrum indicates the presence of both thermal and non-thermal emission, with a best-fit photon index $\Gamma\sim1.2$ and X-ray luminosity $\mathrm{L_X(0.3-8\,keV)}=8.33\times10^{31}\,\mathrm{erg/s}$, typical of redback pulsars \citep{Linares+14}. \citet{Lu+11} searched for the optical counterpart to the X-ray source using Hubble Space Telescope (HST) observations. They found nine possible candidates within the $95\%$ error circle of the X-ray source position (see Fig.~8 and Table~3 in \citealt{Lu+11}). Based on the colour--magnitude diagram (CMD) analysis, they selected counterpart b as the most promising, and classified the system as a cataclysmic variable, while the remaining eight candidates were interpreted as chance alignments.

In this work we present the identification and characterisation of the companion star to the redback MSP M92A. The paper is organised as follows. In Sect.~\ref{sec-2} we present the datasets and data reduction; in Sect.~\ref{sec-3.1} we discuss the identification and physical properties of the optical companion star to M92A; and in Sect.~\ref{sec-3.2} we analyse the light curve of the star. We discuss our results and draw our conclusions in Sect.~\ref{sec-4}.

\section{Datasets and data reduction}
\label{sec-2}
The adopted dataset was obtained using high-resolution and deep observations secured with the HST. All data were acquired using either the ultraviolet--visible (UVIS) channel of the Wide Field Camera 3 (WFC3), or the Wide Field Channel (WFC) of the Advanced Camera for Surveys (ACS), covering a wavelength range from the near-ultraviolet to the optical through seven different filters. The observations were conducted under five different GO programmes. A summary of the dataset is presented in Table~\ref{tab-1}.

The photometric analysis was performed on pre-reduced images with the \texttt{$\_$flc} extension, which are corrected for flat-field, dark current, and charge transfer efficiency (CTE). After correcting for the pixel area map effects, we performed the point spread function (PSF) fitting using DAOPHOT II \citep{Stetson+87}. First, we modelled a spatially varying PSF for each image by using about 200 bright, well-distributed, and isolated stars. The obtained PSF model  was then used to fit all the sources with a flux peak above $4\sigma$ from the background level. We created a reference list of stars by selecting those identified in at least half of the images of the near-UV filters (F225W and F275W) and we then forced the PSF fit to the corresponding location of these sources in all the other images by using DAOPHOT/ALLFRAME \citep{Stetson+94}. This approach, dubbed the UV-route, was proposed and extensively used
to optimise the detection of faint and hot stars in dense and old stellar populations \citep[e.g.][]{Ferraro+97, Ferraro+03b, raso+17, Chen+22, Chen+23, Cadelano+22b}.  This technique mitigates the crowding effects caused by the presence
of giants and main-sequence turn-off (MS-TO) stars, thus sensibly increasing the level of completeness of the observed samples.
The instrumental magnitudes were finally calibrated to the VEGAMAG system using appropriate zero points and transformations for each filter.

As the images from WFC3 and ACS are both affected by significant geometric distortions, we corrected this effect by using the equations provided by \citet{Bellini+09} and \citet{Sirianni+05}. Then, the instrumental positions were transformed to the International Celestial Reference System using 
cross-correlation 
with the Gaia DR3 Catalogue. The transformations resulted in a total astrometric accuracy of 13 mas.

\section{The optical companion to M92A}

\subsection{Identification of the optical counterpart star}
\label{sec-3.1}
\begin{figure}
\centering
    \includegraphics[width=0.23\textwidth]{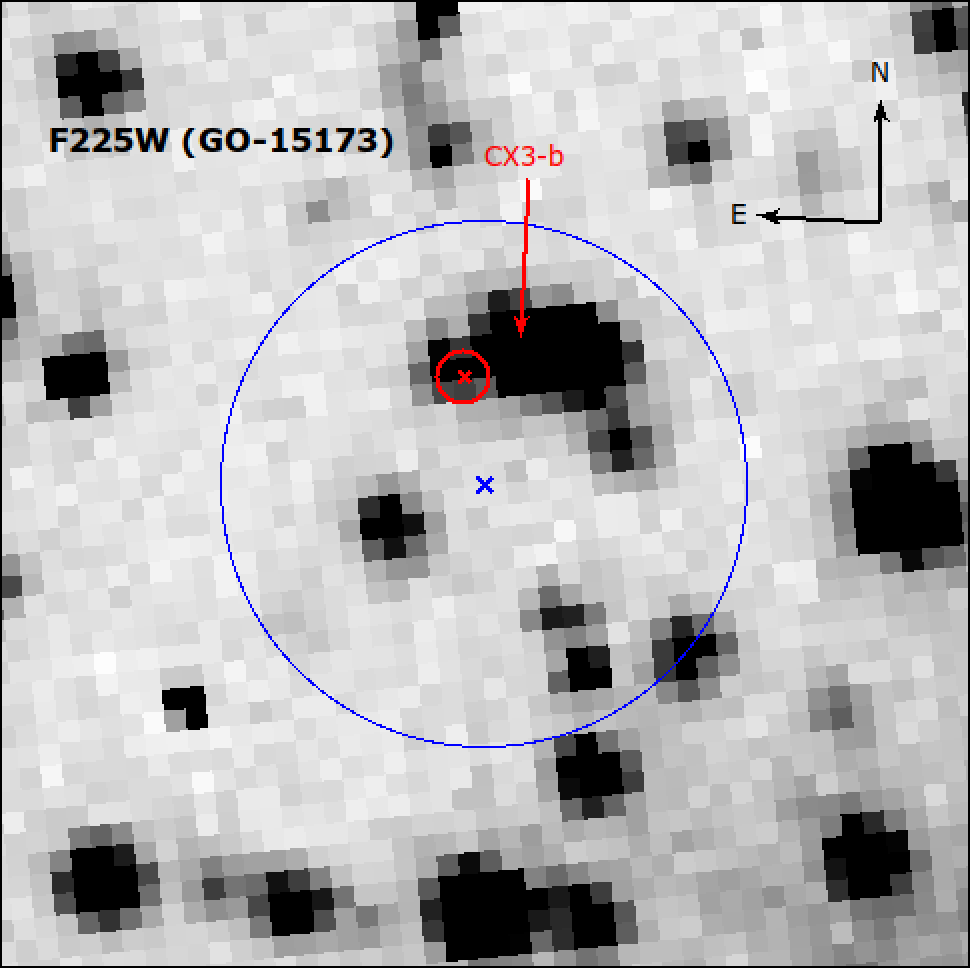}
    \includegraphics[width=0.23\textwidth]{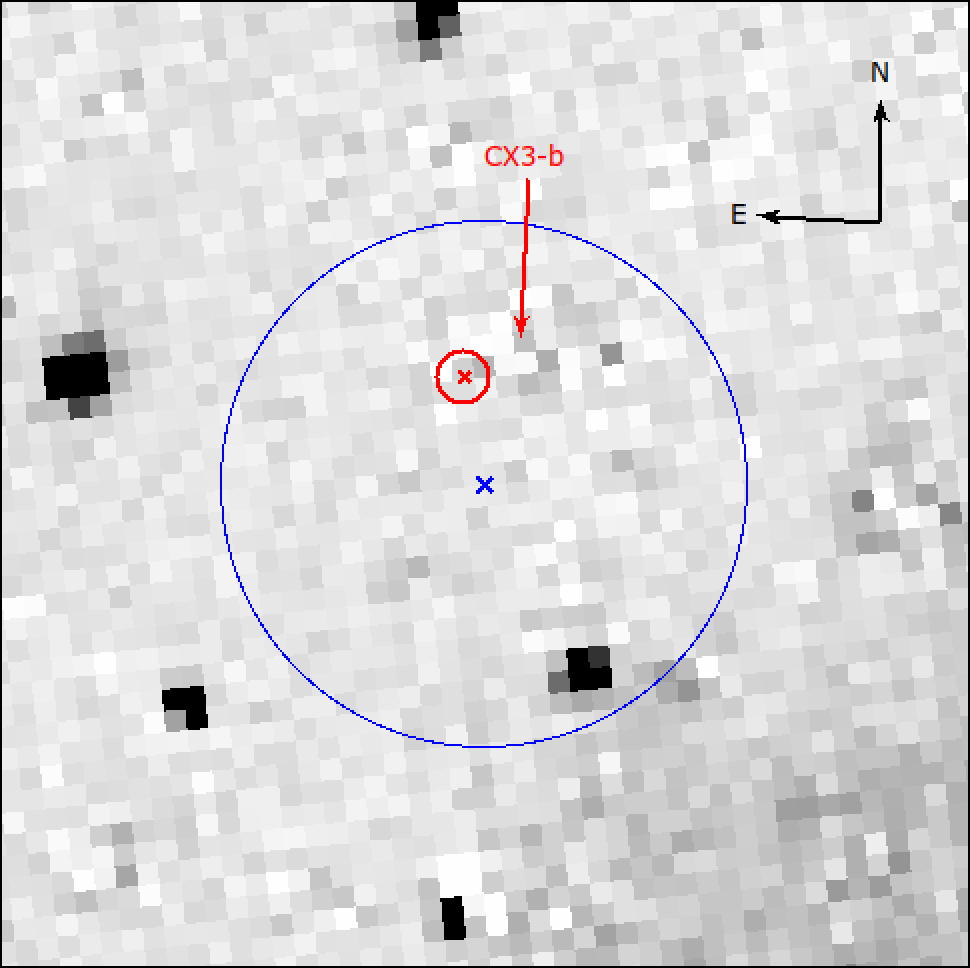}
\caption{
Finding charts of the optical companion star to M92A.  Left panel: Original F225W image of the field around the radio position of M92A. The red cross and red circle mark the radio position of M92A and the corresponding combined optical and radio $\sim 3\sigma$ uncertainty. The companion star identified in this work is clearly visible within this region, nearly coincident with the radio position. For reference, the blue cross and blue circle indicate the nominal X-ray position of CX3 and its $95\%$ confidence error circle from \citet{Lu+11}, while CX3-b, previously discussed by \citet{Lu+11} as a candidate optical counterpart to the X-ray source CX3, is marked with a red arrow.  Right panel: Same as the left panel, but for the PSF-subtracted residual image, demonstrating that nearly all genuine sources are well fitted and subtracted by the PSF model, thus confirming that the photometry of the companion star is not affected by crowding or blending. A few isolated residual spots are due to cosmic rays. All the images have a size of $1.8''\times1.8''$, and the corresponding GO programme numbers are also labelled.
}
\label{fig:FindingChart}
\end{figure}

To identify the optical counterpart to M92A, we carefully examined all the sources detected in the vicinity of the pulsar's radio position. 
As shown in Fig.~\ref{fig:FindingChart}, 
the closest source is a moderately bright star, located at 
$\alpha_{J2000}=17^\mathrm{h}\!:17^\mathrm{m}\!:06^\mathrm{s}\!.499$ and $\delta_{J2000}=+43\degree\!:08'\!:03''\!.48$, 
thus offset by only $0.02''$
from the radio position.
For the sake of clarity,  the candidate counterpart b proposed by \citet{Lu+11} for the X-ray source CX3 (associated with PSR~J1717+4308A by \citealt{Zhao+22}) is also highlighted in Fig.~\ref{fig:FindingChart}.
The latter clearly lies at a larger angular separation ($\sim0.1''$), 
while the former 
 is closer and also displays 
properties that link it to M92A (see below).
Hence, we label the newly identified star as COM-M92A.

\begin{figure*}
    \sidecaption
	\centering
	\includegraphics[width=0.35\textwidth]{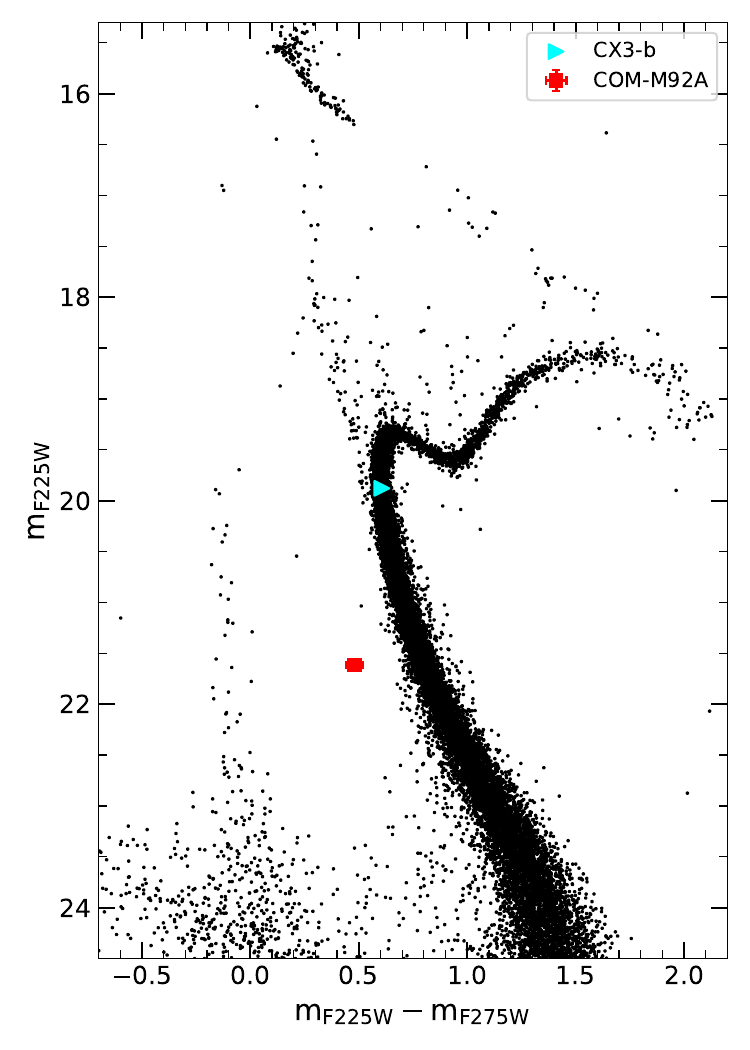}
	\includegraphics[width=0.35\textwidth]{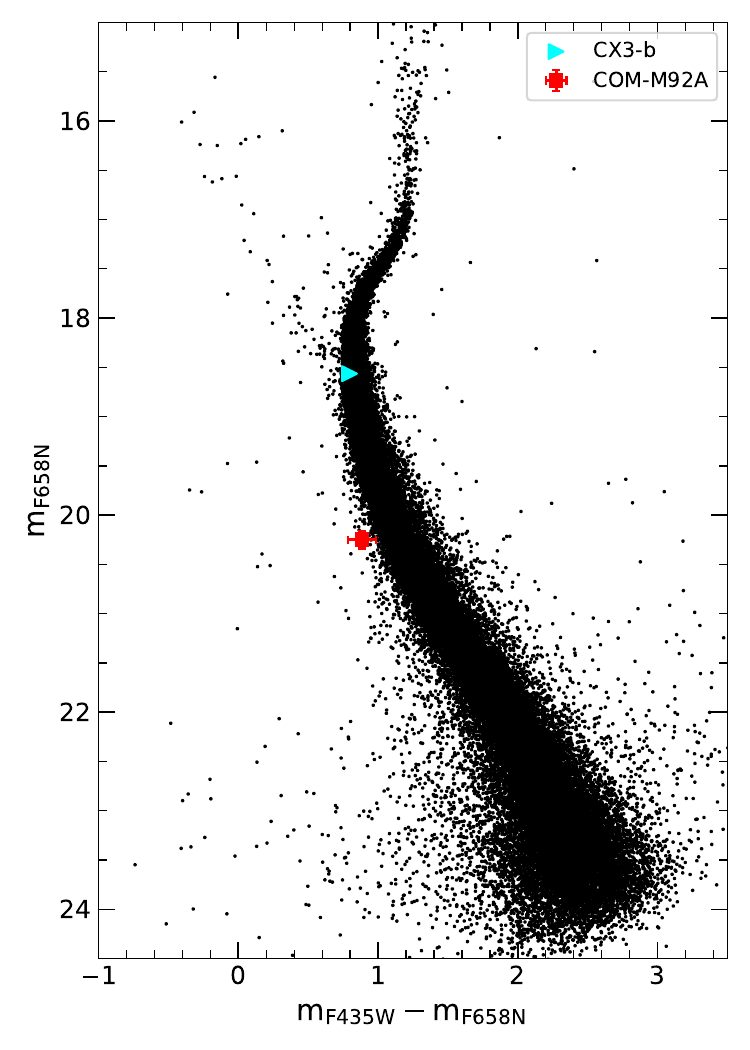}
	\caption{Colour-magnitude diagrams of M92 in ultraviolet and optical filter combinations. In each panel, the red square marks the optical companion to M92A identified in this work, with phase-averaged magnitudes from multi-epoch photometry.
    The cyan triangle corresponds to optical counterpart b of X-ray source CX3 \citep{Lu+11}.}
	\label{fig:cmd}
\end{figure*}

Figure~\ref{fig:cmd} shows two CMDs 
obtained in different filter combinations. The positions of COM-M92A and of the CX3-b counterpart proposed by \citet{Lu+11} as the optical counterpart for the X-ray source are highlighted. 
Interestingly, while the latter occupies CMD positions compatible with being a standard cluster star, the former is located outside the main sequence on the blue side. This offset is particularly evident in the near-UV filter combinations, where COM-M92A is located in the gap region between the main sequence and the white dwarf sequence, suggesting that it could be either an exotic star or a field interloper. 

\begin{figure}
	\centering
	\includegraphics[width=0.45\textwidth]{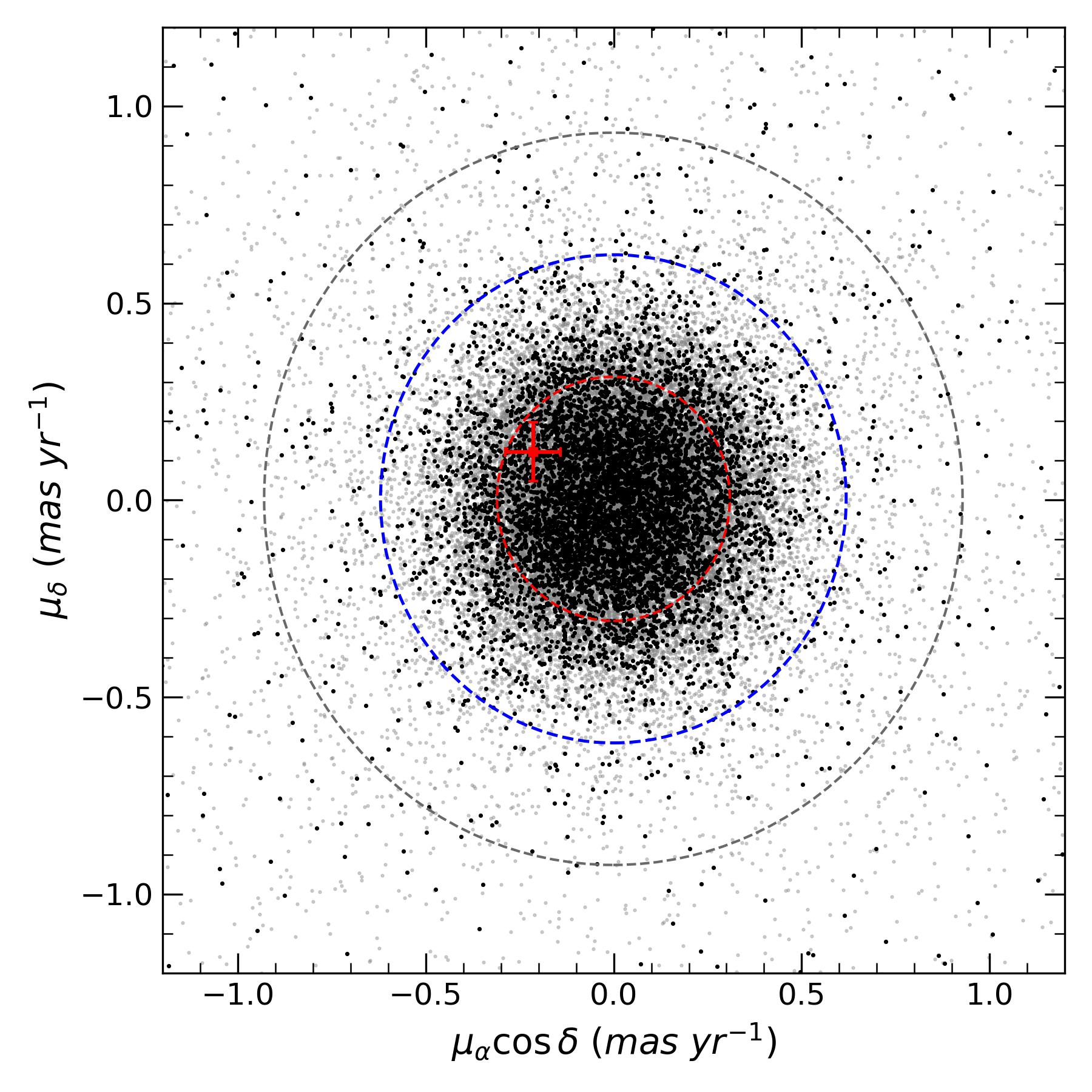}
	\caption{Proper motion vector-point diagram of M92 in the cluster reference frame. All stars are shown as grey dots, while stars within $\pm0.5$~mag of COM-M92A are highlighted in black. The $1\sigma$, $2\sigma$, and $3\sigma$ boundaries derived from the black dot sample are indicated by the red, blue, and grey circles, respectively. The optical companion to M92A is marked by a red square with an error bar.  }
	\label{fig:pm}
\end{figure}

To rule out the second possibility, we analysed the proper motion of this star with respect to that of the other cluster stars. 
We used the photometric and kinematic catalogue of M92 obtained from the HST Atlases of Cluster Kinematics  \citep[HACKS;][]{Libralato+22}. 
The proper motion vector-point diagram of M92 is shown in Fig.~\ref{fig:pm} in the cluster reference frame. Most stars are tightly clustered around the origin, as expected for cluster members, and the position of the investigated candidate companion to M92A is highlighted in red. Remarkably, the relative proper motion of the star 
 \citep[$\mu_\alpha\cos\delta=-0.215\pm0.073$ mas yr$^{-1}$, $\mu_\delta=0.123\pm0.075$ mas yr$^{-1}$;][]{Libralato+22} 
is in excellent agreement with the cluster bulk motion, ruling it out as a field intruder. 
To quantify this agreement, we computed the $1\sigma$, $2\sigma$, and $3\sigma$ boundaries of the proper motion distribution using stars within $\pm0.5$~mag of COM-M92A, and found that COM-M92A falls well within the $1\sigma$ boundary, further confirming its cluster membership.

\begin{figure*}
    \sidecaption
	\centering
	\includegraphics[width=0.7\textwidth]{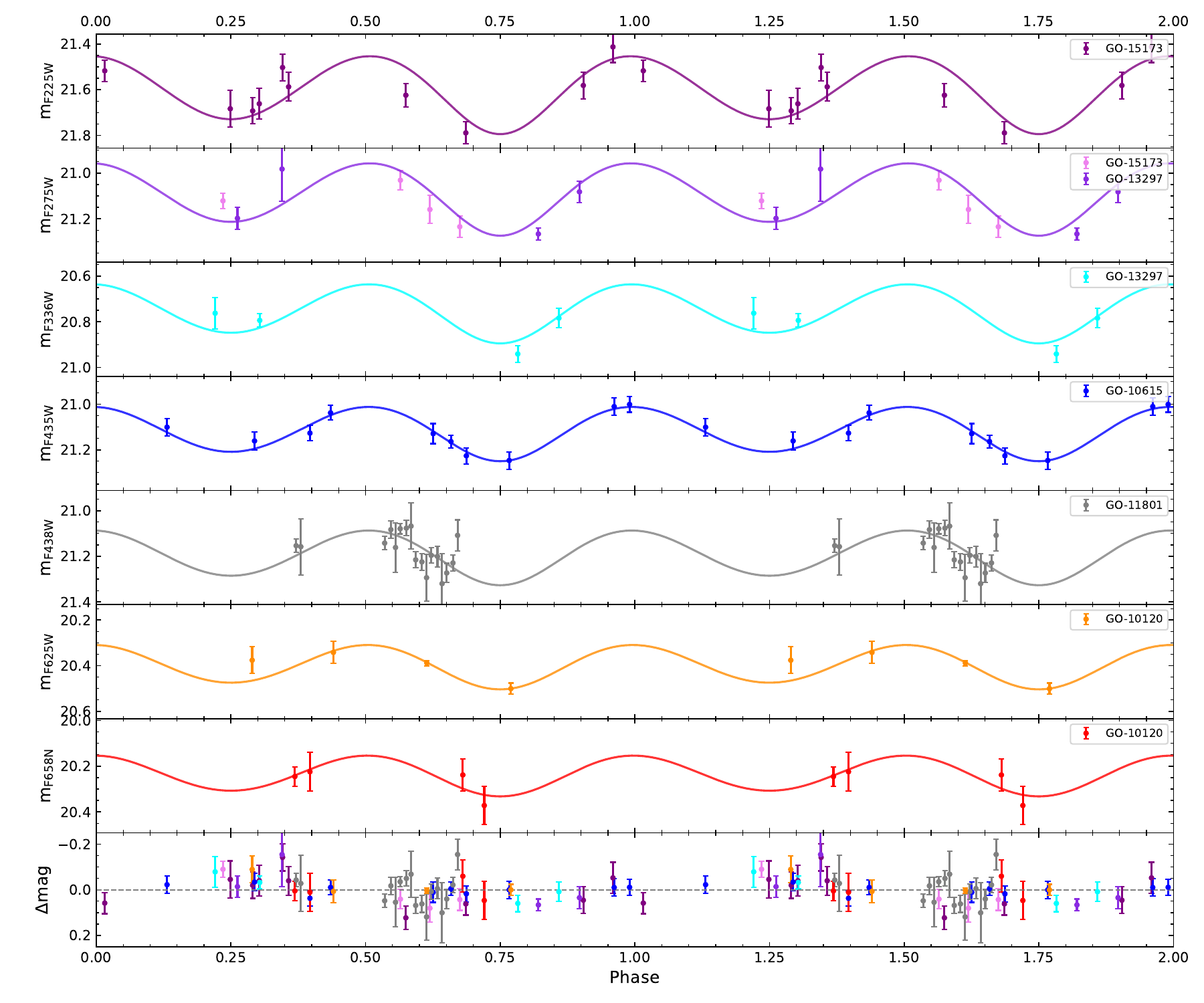}
	\caption{Phase-folded HST light curves of COM-M92A
    in the seven different filters used in this study, overplotted with the best-fit model obtained using the \textsc{Icarus} code (see Section \ref{sec-3.2}). 
    The bottom panel reports all the residuals, colour-coded as in the above panels.
    The orbital phase is defined such that phase zero corresponds to the passage at the ascending node.
    }
	\label{fig:lcsg}
\end{figure*}

\begin{figure*}
    \sidecaption
	\centering
	\includegraphics[width=0.7\textwidth]{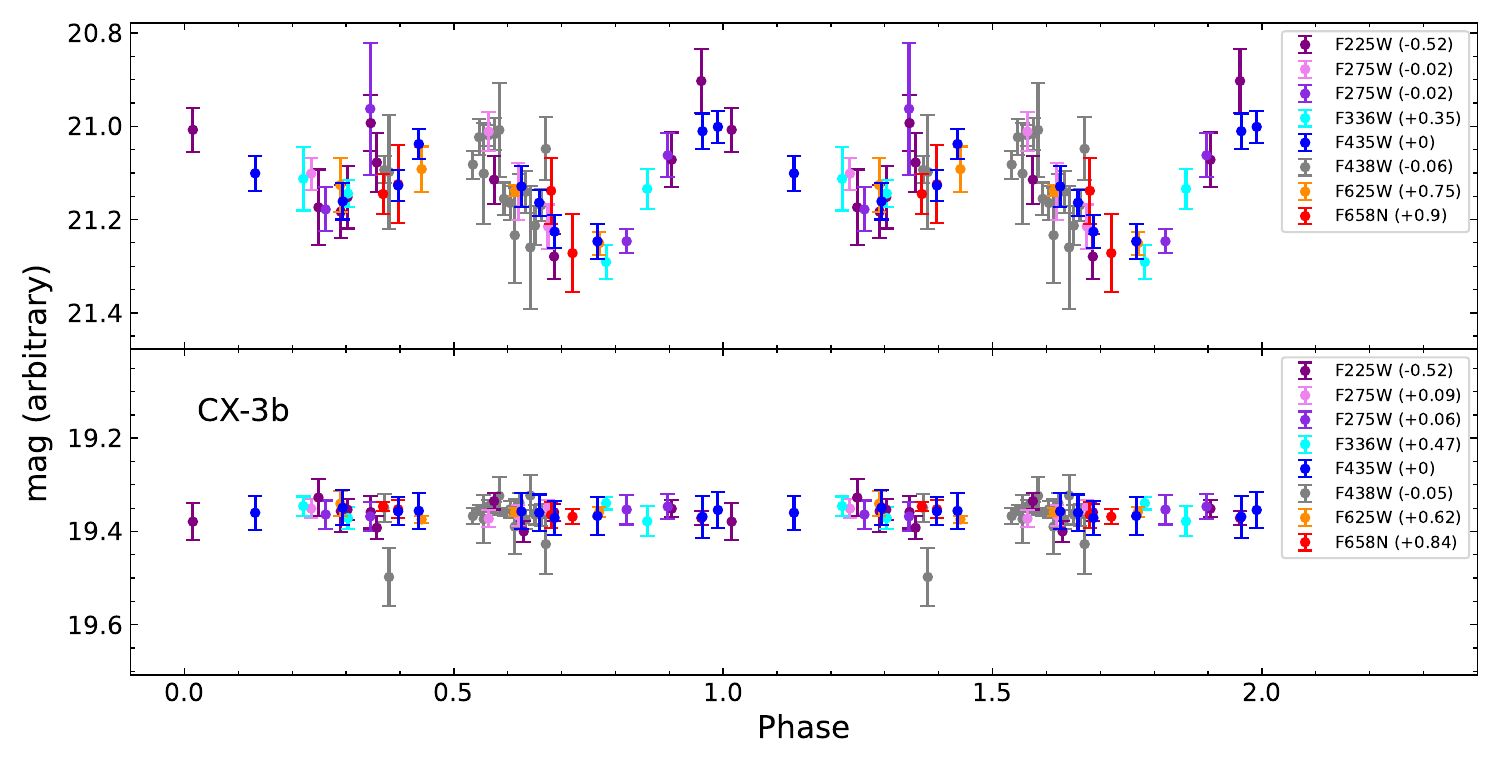}
	\caption{ Upper panel: Global light curve of the optical companion to M92A, 
    obtained from the combination of all the individual light curves shown in Fig.~\ref{fig:lcsg}. The light curve in the F435W filter is chosen as the reference one, and the other light curves are shifted with specific offsets (see legend) to construct a coherent phase-folded profile. 
     Bottom panel: Light curve of CX-3b shown for comparison. CX-3b is clearly a star with no significant variability, confirming that it is not the optical counterpart of a redback system.}
	\label{fig:comlc}
\end{figure*}

We then examined the photometric variability across the available observations of all the sources close to the pulsar's position. We quantified the variability of all sources within the blue circle shown in Fig.~\ref{fig:FindingChart} by computing the reduced chi-squared against a constant-flux null hypothesis, which properly accounts for the photometric uncertainties. COM-M92A yields $\chi^2_\nu = 5.73$ ($p = 5.4 \times 10^{-8}$), while all other sources in the same region have $\chi^2_\nu \le 0.15$, consistent with photometric scatter alone. This confirms that COM-M92A is the only source whose flux variations significantly exceed the measurement errors.
We folded all the magnitudes for each filter using the 
orbital period derived from radio-timing ($P = 0.20087$ days) and the reference epoch ($T_0 \approx 58353.549$) provided by \citet{Pan+20}. The light curves 
obtained in the different filters are shown in Fig.~\ref{fig:lcsg} and combined in Fig.~\ref{fig:comlc}, where a shift in magnitudes has been applied for display purposes. As can be seen, the star shows a magnitude modulation in accordance with the pulsar's orbital period. In particular, it has two maxima, at orbital phases $\phi=0.5$ and $\phi=0.0$, corresponding to the system observed in quadrature, and two minima at $\phi=0.25$ and $\phi=0.75$, corresponding to the pulsar's superior and inferior conjunction, respectively. 
Such a light curve shape, likely due to tidal distortions of the companion due to the pulsar's gravitational pull, has already been observed in the case of other redback companions, such as M28H and NGC6266B \citep{Pallanca+10, Cocozza+08}. 
For comparison purposes, in Fig.~\ref{fig:comlc} we also plot the light curve of CX3-b, which clearly illustrates that this star shows no optical variability.

Based on all the collected evidence in terms of proximity to the radio position, location in the CMD, and magnitude modulation, we conclude that COM-M92A is most likely the optical companion to M92A.

Finally, we checked for a photometric indication of H$\alpha$ emission using the F658N filter. In particular, we examined the position of the stars in the (F435W-F625W, F625W-F658N) colour--colour diagram, which has been shown to be a powerful tool for distinguishing H$\alpha$ emitters among main-sequence stars \citep{Cocozza+08, Pallanca+13, Ettorre+25b}. 
The position of the pulsar companion is compatible with that of other main-sequence stars, suggesting the absence of H$\alpha$ emission. 
In redback systems, H$\alpha$ emission can originate from ongoing 
accretion \citep[see e.g. the transitional MSP M28I in][]{Pallanca+13}, 
from chromospheric activity on the companion star, or from an 
intra-binary shock \citep{Strader+19}. 
The lack of detectable H$\alpha$ emission in COM-M92A is therefore 
consistent with the system being in a rotation-powered, disk-free state.

\subsection{Light curve modelling}
\label{sec-3.2}
We modelled the light curves of COM-M92A using the stellar binary light curve synthesis code \textsc{Icarus} \citep{Breton+12}, with the atmosphere grids generated from the \textsc{atlas9} code \citep{Kurucz+93}. The dataset was converted from Vega apparent magnitudes to flux densities, using the zero-point flux densities provided by the SVO Filter Profile Service.\footnote{\url{https://svo2.cab.inta-csic.es/theory/fps/index.php?&gname=HST&gname2=ACS_WFC}; \url{https://svo2.cab.inta-csic.es/theory/fps/index.php?&gname=HST&gname2=WFC3_UVIS1}}
Since the number of data points per filter differs, to ensure uniform weighting across all filters and avoid any bias towards the filters with the most data points, during the fitting procedure each light curve was resampled by randomly selecting existing data points with replacement within the corresponding filter until a total of 20 data points per filter was obtained. 
To determine the effective temperature of the companion star, we assumed that its base temperature is modified only by the effects of gravity darkening and irradiation from the pulsar wind. To model this, we specified 11 parameters in \textsc{Icarus}. Among these, three parameters were held fixed, while the remaining eight were treated as free parameters in the fitting procedure. We set the orbital period to $\mathrm{P_{orb}} = 0.2008678775(8)$~days \citep{Pan+20}, assumed the system to be tidally locked and so set the co-rotation factor $\omega = 1$, and set the gravity darkening coefficient $\beta = 0.08$ for low-mass companions with a convective envelope \citep{Lucy+67}. The other eight parameters include the semi-major axis of the pulsar $x_1$; the orbital inclination $i$; the projected radial velocity semi-amplitude of the companion $K_2$; the
Roche-lobe filling factor $f = r_{\mathrm{nose}}/r_{L_{1}}$, 
defined as the ratio of the radius of the companion star in the direction of the pulsar to the distance between the companion centre of mass and the inner Lagrangian point; the base temperature 
of the companion prior to irradiation $\mathrm{T_{base}}$; the irradiating temperature $\mathrm{T_{irr}}$, 
which represents the effect of the pulsar heating; the distance $D$; and the extinction in the V band $\mathrm{A_V}$.

We first derived the projected radial velocity semi-amplitude of the pulsar $K_1$, using $x_1$ and $P_\mathrm{orb}$ as done in \cite{Sen+24}. A uniform prior was adopted for $x_1$, using the reported value of $x_1 = 0.398703(2)$~lt-s from \citet{Pan+20} to define the bounds. We then used the derived $K_1$ together with the sampled $K_2$ to compute the mass ratio $q$. We also applied uniform priors between $10$ and $1000$~km $\mathrm{s}^{-1}$ on $K_2$, and between $0$ and $1$ on both $\cos i$ and the filling factor. The prior on the inclination is due to the fact that edge-on systems are more likely to be detected. For both $\mathrm{T_{base}}$ and $\mathrm{T_{irr}}$, we used uniform priors between $1000$ and $15000$~K and $0$ and $10000$~K, respectively. For our last two input parameters, we applied Gaussian priors:\footnote{
For completeness, we also tested a fit with distance and absorption fixed. However, since the derived parameters are consistent within $1\sigma$, we preferred to retain the fit that accounts for the parameter uncertainties and degeneracy.
} for $D$, we used $8.2 \pm 0.6$~kpc \citep{Rees+92} as the distance to the GC M92, while for $\mathrm{A_V}$, we used the colour excess from the \cite{Green+19} dust maps of $\mathrm{E}(g'-r') = 0.02 \pm 0.02$. The final constraint we applied was only to allow models with a NS mass of less than $3 \,\mathrm{M}_{\odot}$.

\begin{figure*}
    \sidecaption
	\centering
	\includegraphics[width=0.7\textwidth]{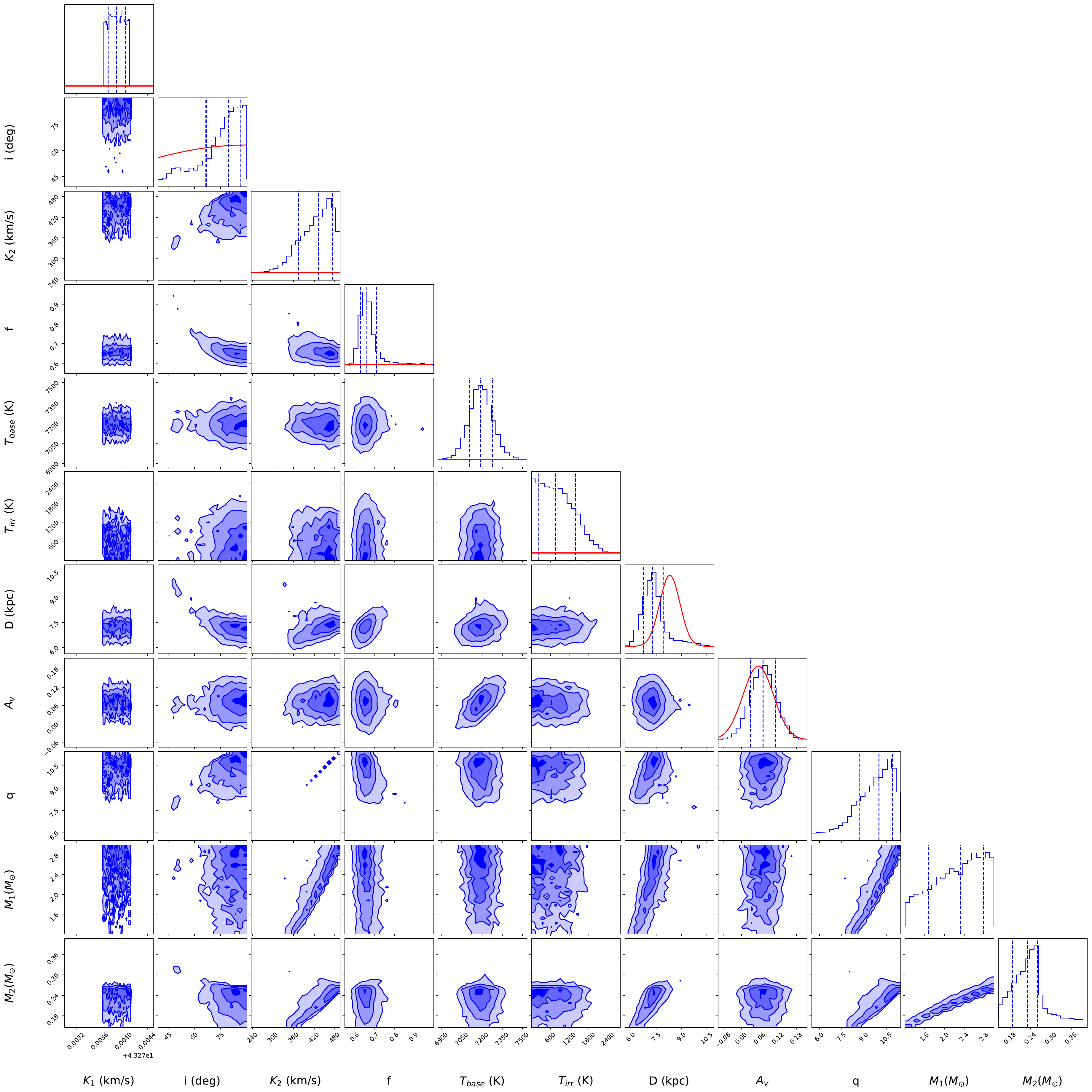}
	\caption{Corner plot showing the posterior distributions and parameter correlations from the \textsc{dynesty} fit of the eight free parameters  in the \textsc{Icarus} model. The marginalised distributions for inclination $i$, radial velocity semi-amplitude $K_2$, filling factor $f$, base temperature $T_\mathrm{base}$, irradiation temperature $T_\mathrm{irr}$, distance $D$, and extinction $A_V$ are shown along the diagonal. Derived parameters such as the mass ratio q, the pulsar mass $M_1$, and the companion mass $M_2$ are also included. 
    Prior distributions are shown as red curves. Notably, the extinction $A_V$ closely follows its prior distribution, while $D$ remains consistent with its prior within $1\sigma$. In contrast, $K_2$ and $i$ are less well constrained, contributing to the broader posterior on $M_1$.}
	\label{fig:corner}
\end{figure*}

Using the nested sampler \textsc{dynesty} \citep{Speagle+20, Skilling+04, Skilling+06, Feroz+09, Koposov+24} we fit for the eight parameters. The results for our best-fit model are reported in Table~\ref{tab:lcmodelresults}, and we show the light curves  (and residuals) for this fit in Fig.~\ref{fig:lcsg}. In Fig.~\ref{fig:corner} we show the corner plot, where we note that the distribution of $\mathrm{A_V}$ closely follows its prior and $D$ is consistent with its prior to $1\sigma$. 
The value of $K_1$ is well constrained from the precise radio constraint of $x_1$. From this, the derived companion mass is well constrained. In contrast, both $i$ and $K_2$ are less well constrained, resulting in large uncertainties on the mass of the pulsar,  $\mathrm{M}_{1} = 2.3_{-0.6}^{+0.5} \,\mathrm{M}_{\odot}$.

Based on the fitting results from \textsc{dynesty}, the companion star to M92A is predicted to be a hot  ({$\mathrm{T} = 7054^{+91}_{-88}$~K}) 
perturbed star with a mass of $\mathrm{M_{COM}} = 0.23^{+0.03}_{-0.04}\,\mathrm{M}_{\odot}$, orbiting a NS with mass $\mathrm{M_{NS}} = 2.3^{+0.5}_{-0.6}\,\mathrm{M}_{\odot}$. The relatively high orbital inclination of $i = 79^{+7}_{-13}$ degrees is consistent with the observed optical modulation and the presence of radio eclipses.

\begin{table}
\renewcommand{\arraystretch}{1.3} 
\centering
    \begin{tabular}{cc}
        \toprule
        Fitted & \\
        \midrule
        $i$ (deg) & $79_{-13}^{+7}$ \\
        $K_2$ (km/s) & $433_{-58}^{+40}$ \\
        $f$ & $0.66_{-0.03}^{+0.05}$ \\
        $\mathrm{T_{base}}$ (K) & $7191_{-85}^{+88}$ \\
        $\mathrm{T_{irr}}$ (K) & $754_{-525}^{+620}$ \\
        $D$ (kpc) & $7.3_{-0.5}^{+0.6}$ \\
        $\mathrm{A_V}$ (mag) & $0.07 \pm 0.04$ \\
        \midrule
        Derived & \\
        \midrule
        q & $10.0_{-1.3}^{+0.9}$ \\
        $\mathrm{M}_1 (\mathrm{M}_{\odot})$ & $2.3_{-0.6}^{+0.5}$ \\
        $\mathrm{M}_2 (\mathrm{M}_{\odot})$ & $0.23_{-0.04}^{+0.03}$ \\
        $\mathrm{T_{sup}}$ (K) & $7051_{-85}^{+88}$ \\
        $\mathrm{T_{inf}}$ (K) & $7057_{-85}^{+88}$ \\
        $\mathrm{L_{irr}}$ ($10^{30}$ erg/s) & $3.2 \pm 1.3$ \\
        \midrule
        $\chi^2_\nu$ & 1.18 \\
        \bottomrule
    \end{tabular}
    \caption{ Best-fit linked model results, with the 50th percentile value reported with the 16th and 84th percentiles as uncertainties. The fitted parameters are reported at the top of the table and the derived parameters in the middle. These are the mass ratio $q$, reported in the form $\mathrm{M_{1}/M_{2}}$; the pulsar mass $\mathrm{M_{1}}$ and the companion mass $\mathrm{M_{2}}$; the companion temperature at the superior ($\mathrm{T_{sup}}$) and inferior ($\mathrm{T_{inf}}$) conjunction; and the irradiation luminosity $\mathrm{L_{irr}}$. The bottom section reports the reduced $\chi^2$. 
 }
    \label{tab:lcmodelresults}
\end{table}

\section{Discussion and conclusions}
\label{sec-4}
We have identified and characterised the optical companion to the redback millisecond pulsar in M92, PSR J1717+4308A, using deep, multi-epoch HST observations. The star proposed as the optical counterpart lies at only $0.02''$ 
from the updated radio timing position. 
Moreover, from the analysis of its proper motion, we confirm that the companion star is a cluster member. The absence of detectable photometric H$\alpha$ emission strongly suggests that no active accretion is occurring and that M92A is observed in a purely rotation-powered state. However, the strongest confirmation of the physical association with the NS comes from the star’s out-of-sequence position in the CMD and, most importantly, from the detection of photometric variability with a periodicity fully consistent with the 0.2-day orbital period derived from radio timing.

The light curve, which shows two minima at the pulsar's superior and inferior conjunctions, suggests that the companion is tidally deformed by the pulsar's gravitational field. As confirmed by \citet{Turchetta+23}, approximately half of the known redback population displays similar double-minimum light curves in the absence of strong irradiation. However, this morphology is not universal. For instance, \citet{Turchetta+25} observed a single-minimum modulation for the optical counterpart to PSR~J2055+1545 \citep{Lewis+23}, attributed to mild heating by the pulsar wind, just sufficient to exceed the underlying ellipsoidal modulation.

\citet{Turchetta+23} showed that the presence or absence of irradiation in such systems is well captured by the pulsar spin-down to the companion flux ratio $f_\mathrm{sd}$, which quantifies the strength of pulsar wind irradiation. A transition from ellipsoidal-dominated to irradiation-dominated regimes occurs at $f_\mathrm{sd} \sim 2\text{--}4$. This parameter depends on the orbital period $\mathrm{P_{orb}}$, the companion base temperature $\mathrm{T_{base}}$, and the pulsar's spin-down luminosity $L_\mathrm{sd}$, and can be used to predict the expected light curve behaviour under the assumption of isotropic pulsar wind emission. 
For M92A, we computed $f_\mathrm{sd}$ using Eq.~4 from \citet{Turchetta+23}, adopting the base temperature from our best-fit model and calculating $L_\mathrm{sd}$ starting from the radio timing solutions. We obtained $f_\mathrm{sd} \approx 2.71$, which places the system near the transition threshold. Interestingly, while M92A exhibits very low irradiation, consistent with other non-irradiated redback systems, our light curve fitting results, similar to \citet{Sen+24}, indicate that including a small amount of irradiation improves the fit, 
 with non-irradiated models yielding a $\sim 0.1$ higher reduced chi-squared.
This result can be understood by looking at the posterior distributions shown in Fig.~\ref{fig:corner}. The corner plot shows that the best-fit solution for $\mathrm{T_{irr}}$ allows   values higher than zero, although with large uncertainties, and that the likelihood distribution clearly peaks at values close to zero. This indicates that only a very small level of irradiation is statistically preferred and this behaviour is consistent with the low value of $f_\mathrm{sd}$ estimated for the system, which suggests that irradiation effects, if present, should be weak.

Moreover, the light curve modelling performed with \textsc{Icarus} provides further quantitative evidence supporting the redback nature of the system. The best-fit solution yields a high orbital inclination ($i = 79^{+7}_{-13}$ degrees) and a 
mass ratio $\rm{q}=10.0_{-1.3}^{+0.9}$, which are consistent with the typical values found for redback systems (black widows generally have higher q; see e.g. \citealt{Strader+19}). The individual resulting masses are $\mathrm{M_{COM}} = 0.23^{+0.03}_{-0.04}\,\mathrm{M}_{\odot}$ and 
$\mathrm{M_{NS}} = 2.3^{+0.5}_{-0.6}\,\mathrm{M}_{\odot}$.
Finally, a remarkable feature of M92A is the temperature of its optical companion. With $\mathrm{T} = 7054^{+91}_{-88}$~K, this system ranks among the hottest known redbacks, second only to PSR~J1431$-$4715, which has $\mathrm{T_{sup}} \sim 7500$~K and $\mathrm{T_{inf}} \sim 7400$~K \citep{deMartino+24}. 
The unusually high temperature of the companion makes this system an outlier within the redback population, together with PSR~J1431$-$4715, as the majority of known redbacks have companion temperatures below $\sim 6000$ K,
although unusually hot or blue companions have been noted in other spider systems \citep{Schroeder+14}.
In contrast, the companion in this system shows a temperature typical of an F-type star, a property shared only with PSR~J1431$-$4715 among currently known redbacks. This may indicate an atypical evolutionary history. This possibility is particularly plausible given that the system resides in a GC, where dynamical interactions can favour the production of non-standard binary configurations compared to pulsars belonging to the field.
Finally, it is worth noting that inspection of the residual panels in Fig.~\ref{fig:lcsg} suggests that an unrecovered phase-dependent structure may still be present. This feature is likely related to second-order effects in the light curve that, because of the limited statistics, cannot be reliably modelled with the current dataset.

Taken together, the properties of M92A make it a twin of M28H and also make it similar to other redbacks identified in GCs, such as NGC~6397A and NGC~6266B 
\citep{Ferraro+01b,Cocozza+08,Pallanca+10}, and 
in the Galactic field \citep[e.g. PSR~J1431$-$4715;][]{deMartino+24}.
At the same time, the combination of its high effective temperature, its high mass ratio, and the possibility of a relatively massive NS make this system particularly noteworthy. 

These characteristics offer useful clues for interpreting the evolution of compact binaries in dense stellar environments, especially with regard to the role of irradiation and the extent of tidal deformation.
The present analysis also confirms the power of high-resolution, multi-band, and, most importantly, multi-epoch images in revealing and characterising compact binaries in crowded cluster regions. The inclusion of M92A in the small but growing set of redback systems hosted by GCs provides fresh insight into the evolutionary channels leading to the formation of MSPs in such environments.

\begin{acknowledgements}
This work has been
funded using resources from the INAF Large Grant 2022 “GCjewels” (P.I. Andrea Possenti) approved with the Presidential Decree 30/2022. 
GE and ED acknowledge financial support from the INAF Data analysis Research Grant (PI E. Dalessandro) of the “Bando Astrofisica Fondamentale 2024”.
The research activities described in this paper were carried out with contribution of the Next Generation EU funds within the National Recovery and Resilience Plan (PNRR), Mission 4 - Education and Research, Component 2 - From Research to Business (M4C2), Investment Line 3.1 - Strengthening and creation of Research Infrastructures, Project IR0000034 – ``STILES - Strengthening the Italian Leadership in ELT and SKA''. JC acknowledges support from the National Science Foundation of China (grant No. 12503036) and the China Postdoctoral Science Foundation (grant No. 2024M760242). BS, DM and ML acknowledge funding from the European Research Council (ERC) under the European Union’s Horizon 2020 research and innovation programme (grant agreement No. 101002352). 
\end{acknowledgements}

\appendix
\section{Dataset summary}
\label{app:table1}
\begin{table}[htp]
\renewcommand{\arraystretch}{1.2} 
\caption{Summary of the multi-epoch HST dataset used in this work.}
\label{tab-1}
\begin{tabular}{lccc} 
\hline \hline
	Proposal ID		& Observation date			& Filter  		&   Exposure time 	\\ \hline
	\multicolumn{4}{c}{WFC3}   \\ \hline
GO-15173 & 2019 June 28      & F225W	 & 	 800s$\times$2, 805s$\times$3, 815s, 820s, 835s$\times$4 \\
         &                   & F275W	    &	800s	$\times$2, 815s, 825s\\
GO-13297 & 2014 August 03    & F275W	 &	 707s$\times$2, 819s$\times$2	\\		
         &                   & F336W	    &	707s$\times$2, 819s$\times$2	\\ 		
GO-11801 & 2009 November 11	 & F438W	 &	  10s$\times$6, 110s$\times$11	\\ \hline		
	\multicolumn{4}{c}{ACS}   \\ \hline
GO-10615 & 2006 January 09	 & F435W	 &	300s, 340s$\times$14	\\ 
GO-10120 & 2004 August 07    & F625W	 &	10s, 120s$\times$3		\\		
         &                   & F658N		&	350s$\times$2,555s$\times$2	\\ \hline
\end{tabular}
\tablefoot{The rightmost column lists the exposure time in seconds for each image.}

\end{table}

\end{document}